\documentclass[prl,tighten,article,twocolumn,epsfig,graphics,showpacs,mathbbm]{revtex4}

\usepackage{amsmath,amsfonts,amssymb,graphics,graphicx,epsfig,color,times,bbm}

\begin{document}
\bibliographystyle{apsrev}

\newcommand{\R}{\mathbbm{R}}
\newcommand{\rr}{\mathbbm{R}}
\newcommand{\nn}{\mathbbm{N}}
\newcommand{\cc}{\mathbbm{C}}
\newcommand{\zz}{\mathbbm{Z}}

\newcommand{\ii}{\mathbbm{1}}

\newcommand{\id}{\mathbbm{1}}

\newcommand{\tr}[1]{{\rm tr}\left[#1\right]}
\newcommand{\gr}[1]{\boldsymbol{#1}}
\newcommand{\be}{\begin{equation}}
\newcommand{\ee}{\end{equation}}
\newcommand{\bea}{\begin{eqnarray}}
\newcommand{\eea}{\end{eqnarray}}
\newcommand{\St}{{\cal S}}
\newcommand{\Ad}{{\rm Ad}}

\newcommand{\ket}[1]{|#1\rangle}
\newcommand{\bra}[1]{\langle#1|}
\newcommand{\braket}[2]{\langle #1|#2\rangle}
\newcommand{\avr}[1]{\langle#1\rangle}
\newcommand{\G}{{\cal G}}
\newcommand{\eq}[1]{Eq.\ (\ref{#1})}
\newcommand{\ineq}[1]{Ineq.\ (\ref{#1})}
\newcommand{\sirsection}[1]{\section{\large \sf \textbf{#1}}}
\newcommand{\sirsubsection}[1]{\subsection{\normalsize \sf \textbf{#1}}}
\newcommand{\ack}{\subsection*{\normalsize \sf \textbf{Acknowledgements}}}
\newcommand{\front}[5]{\title{\sf \textbf{\Large #1}}
\author{#2 \vspace*{.4cm}\\
\footnotesize #3}
\date{\footnotesize \sf \begin{quote}
\hspace*{.2cm}#4 \end{quote} #5} \maketitle}
\newcommand{\eg}{\emph{e.g.}~}

\newcommand{\proofend}{\hfill\fbox\\\smallskip }

%---------------------------------------------------------------------------

\newtheorem{theorem}{Theorem}
\newtheorem{proposition}{Proposition}

\newtheorem{lemma}{Lemma}
\newtheorem{definition}{Definition}
\newtheorem{corollary}{Corollary}
\newtheorem{example}{Example}
\newtheorem{remark}{Remark}
\newtheorem{problem}{Problem}

\newcommand{\proof}[1]{{\it Proof.} #1 $\proofend$}

\title{Quantitative entanglement witnesses}

\author{J.\ Eisert, F.G.S.L.\ Brand\~ ao, 
and K.M.R.\  Audenaert}

\affiliation{
Institute for Mathematical Sciences, 
Imperial College London,
London SW7 2PE, United Kingdom\\
QOLS, Blackett Laboratory, 
Imperial College London,
London SW7 2BW, United Kingdom}

\date{\today}

\begin{abstract}
Entanglement witnesses provide tools to detect entanglement
in experimental situations without the need of having
full tomographic knowledge about the state. If one 
estimates in an experiment an expectation value smaller
than zero, one can directly infer that the state has been
entangled, or specifically multi-partite entangled, in the 
first place. In this article, we emphasize that
all these tests -- based on the very same data --
give rise to quantitative estimates in terms of
entanglement measures: ``If a test is strongly violated,
one can also infer that the state was quantitatively
very much entangled''. We consider various
measures of entanglement, including the negativity,
the entanglement of formation, and the robustness
of entanglement, in the bipartite and multipartite
setting. As examples, we discuss several experiments
in the context of 
quantum state preparation that have recently been
performed.
\end{abstract}

 
\maketitle

\section{Introduction}

Entanglement witnesses have proven
tremendously helpful in the 
experimental characterization
of entanglement in composite quantum
systems 
[1--13].
They are observables from the expectation
values of which one can argue whether a prepared
state is indeed entangled: 
whenever its expectation value takes  a value
smaller than zero, then one can unambiguously draw
the conclusion that the state has been entangled
in a particular fashion [1--4]:
the entanglement has then 
been ``witnessed''. This approach seems particularly
feasible or helpful in situations where one would like
to avoid to collect sufficient data to arrive at
full tomographic knowledge. Specifically in multi-partite
settings when detecting multi-particle entanglement
this can be costly. 
Also in instances one can tolerate larger errors
when estimating entanglement witnesses 
compared to
the procedure where one first estimates the full 
state.

Originally, such a 
test for entanglement was thought to give rise to 
an answer to a ``yes-no-question'': the state is entangled 
or it is not.
Yet, in this way, one does not make use of valuable 
information that one has collected anyway. Actually, one 
has implicitly recorded 
data that are sufficient to make
a {\it quantitative} statement: if a test is very much violated
-- so delivers a value much smaller than zero -- then one can 
infer that in quantitative terms, the state was highly entangled.
This quantitative statement is then meant in terms of some
measure of entanglement. 
This is very useful information: One does 
not only know that the the specific entanglement
property is contained in the state. But one can also give
an answer to the question how useful a given state is,
say, to perform a certain task of quantum information.

This article emphasizes this fact, and 
advocates the paradigm of quantitative tests based on data
from measuring witness operators \cite{NoteThat}.
Needless to say, one should under all circumstances
only make use of the data that have in fact been acquired in an experiment, and
avoid hidden assumptions concerning the nature of the involved states. 
But then, in turn, one should make use of the full information
that can in fact be extracted from the measurement data, including 
quantitative assessments.

\section{Paradigm of quantitative tests}

The paradigm we describe is 
the following: imagine one has collected data from 
a measurement of an {\it entanglement witness},
or  a collection thereof. What is the worst case scenario
one could have had, concerning the degree of entanglement?
Certainly, one should
provide conservative estimates in this context. 
This is typically the practically most relevant question: 
one has prepared a state, and wants to know to what
degree one has succeeded in doing so. This test should
make use of a minimal possible number of data, or
measurement settings, certainly
less than full tomography. So we aim for 
answers to\\

\medskip

\begin{minipage}{.05\columnwidth}
$\,$
\end{minipage}
\begin{minipage}{.8\columnwidth}
	{\it ``Given  measurement
	data from measuring an entanglement witness,
	which one is quantitatively the least
	entangled state 
	consistent with the data?''}\\
\end{minipage}

\medskip

\begin{figure}[!h]
  \begin{center}
      \leavevmode
\resizebox{6.2 cm}{!}{\includegraphics{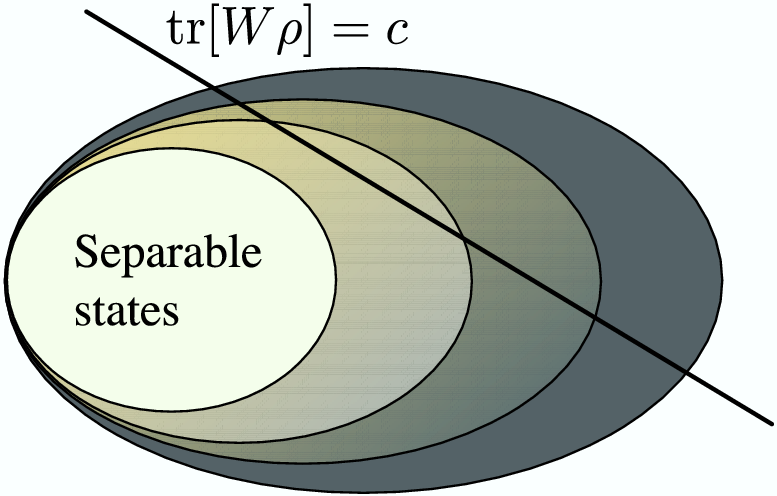}}
\end{center}
\caption{Schematic representation
of state space. The set of separable states is depicted
as the white region. The straight line represents an
experimental test, so the hyperplane characterized by
an entanglement witness $W$ and its expectation
value $\text{tr}[\rho W]=c$.
Then, one encounters
a hierarchy of convex sets of states with increasing degree
of entanglement, as quantified by any
convex entanglement monotone.}
\label{Sep}
\end{figure}

\noindent 
This translates to an {\it optimization problem}: for a given
measure of bi-partite or multi-partite entanglement $E$,
we aim at finding the solution of
\begin{eqnarray}
	\min && E(\rho) 
	\\
	 \text{subject to}  && \rho\,\,\text{ consistent with the
	 data},\nonumber
\end{eqnarray}
or at least get reasonably good lower bounds. The general
spirit of this paper will be to assume nothing more than the 
partial information provided by expectation values of 
entanglement witnesses. Based on this information, we aim
at finding good bounds to entanglement measures. We also
comment on the tightness of these bounds. In fact, the
provided strategies often give rise to the best (tightest)
possible bounds based on this partial information. The ``true
state'' of the system 
is not assumed to be known, or it is not even assumed
that it could in principle be measured, as full 
quantum tomography may be inaccessible. 
In turn, the {\it optimization of entanglement witnesses}, 
so the construction of tangent hyperplanes 
[11, 12, 16--19]
is an interesting (and computationally provably
hard) problem in its own right, which we will not touch upon
here. Any known findings in this field can however immediately
be applied to our setting, in that the entanglement witness
that is most violated will give rise to the best bound.
We hence take the
entanglement witness as such and the corresponding data
for granted, and will provide good bounds for 
entanglement measures 
based on them. This is actually the situation 
one faces 
when interpreting experimental data.

There is a body of work somewhat similar in spirit in the
literature. The need for
conservative estimates, so for minimizing the degree of entanglement
in the context of a Jaynes statistical inference scheme consistent
with the data was already 
noted in the early work Ref.\ \cite{Jaynes}.
Also, conceptually, this is related to a connection of violations
of Bell inequalities to entanglement measures \cite{FM},
and to quantum state estimation as in Ref.\ \cite{Blume}. 

In this work we consider the bi-partite and multi-partite setting.
The system can hence be thought of consisting of a
number of subsystems, such that the 
Hilbert space is given by ${\cal H}= \cc^{d_1}\otimes
\dots\otimes \cc^{d_N}$. 
We assume that we have collected data
that we can estimate $\{c_1,\dots,c_n\}$
based on a number of entanglement witnesses
$W_1,\dots,W_n$, meaning that
\begin{equation}
	\langle W_i\rangle= 
	\text{tr}[W_i \rho]=c_i
\end{equation}
for $i=1,\dots,n$, for example, for a single entanglement witness,
$n=1$. {\it Witness} $W$ 
means in the bi-partite setting,
$N=2$, that for all {\it separable states}
\begin{equation}
	\rho= \sum_i p_i \rho_i^{(1)}\otimes \rho_i^{(2)}
\end{equation}	
on ${\cal H}= \cc^{d_1}\otimes \cc^{d_2}$ 
we have that [1--4]
\begin{equation}
	\text{tr}[W \rho]\geq 0, 
\end{equation}	
and at least for a 
single entangled state $\rho$, one finds that 
\begin{equation}
	\text{tr}[W \rho]<0.
\end{equation}	
This is very intuitive: the separable states form
a convex set, and the witness defines a hyperplane in 
state space that separates the separable states, see Fig.\
\ref{Sep}. In the same way, one can define entanglement witnesses
for the various classes of {\it multi-particle entanglement},
in a setting with Hilbert space
${\cal H}= \cc^{d_1}\otimes \dots\otimes \cc^{d_N}$.
For witnesses in infinite-dimensional systems and
relationships to entanglement measures, see
Refs.\ [23--26].
In this paper, we refrain from introducing these
multi-partite entanglement classes, and refer for that
to Refs.\ [27--30].

We should mention at this point that if one allows
for witnesses taking several identically prepared
specimens into account, one can often improve
the bounds to entanglement measures. On the
positive side, this gives rise to sharper or tight 
bounds, often making use of few different
types of measurements, or indeed
even single ones 
[31--35].
On the negative side, one needs to implement collective
operations, either with quantum networks, or in 
optical settings, with joint operations involving 
bringing together independent sources at
beam splitters. Nonetheless, the first 
experimental measurements
of a two-copy witness for arbitrary two qubit pure states was 
recently reported \cite{Walborn}. 
Although we make the presented ideas explicit for the
most frequently applied approach of
measurements on individual specimens, it should
be noted that many of the presented ideas
are also applicable to this case of collective
operations.

\section{Implications to entanglement measures}

Subsequently, we will present a framework of 
quantitative tests, and discuss a number of 
bounds for different entanglement measures. 
We will also discuss several examples, taken from
the bi-partite and multi-partite context. The concept
of the conjugate function will play a central role here.

\subsection{Negativity}

In this first part we consider  bi-partite {\it splits}
of our system: so the system is either naturally bi-partite,
or we group the subsystems into two parts, with joint
Hilbert space ${\cal H}=\cc^{d_1}\otimes \cc^{d_2}$,
with state space ${\cal S}({\cal H})$. The  
{\it negativity} 
is a measure of entanglement 
defined as
\begin{equation}
	E_N(\rho)=  \|\rho^\Gamma\|_1-1 ,
\end{equation}	
in terms of the trace-norm $\|A\|_1 = \text{tr}|A|$.
$\rho^\Gamma$ denotes the partial transpose of 
$\rho$.
The negativity 
has been  introduced in Ref.\  \cite{Neg}, compare also
Ref.\ \cite{Compare}, and independently
shown to be an entanglement
monotone in Refs.\ \cite{Monotone,VidalWerner}. 
The logarithmic version $\log_2\|\rho^\Gamma\|_1$ of the
negativity is also
an entanglement monotone \cite{Plenio}, 
and a useful upper bound to
the distillable entanglement 
\cite{VidalWerner,Cost}. For this measure of entanglement, 
we will indeed
find very simple, yet tight and useful bounds.

What we are interested in here is the minimally entangled
state consistent with what has been measured.
So we seek the solution of 
\begin{eqnarray}
	E_{N,\text{min}}=\inf && \|\rho^\Gamma\|_1-1\\
	 \text{subject to}  && \text{tr}[\rho W_i] = c_i,\nonumber\\
	&& \rho\geq 0,\nonumber\\
	&& \text{tr}[\rho]=1,\nonumber
\end{eqnarray}
which is the desired quantity. 
Now, this can also be written as
\begin{eqnarray}
	E_{N,\text{min}}=\inf && \max \text{tr}[P  \rho^\Gamma]-1
	\\
	 \text{subject to }  && \|P\|_\infty=1,\nonumber\\
	 && \text{tr}[\rho W_i] = c_i,\nonumber\\
	&& \rho\geq 0,\nonumber\\
	&& \text{tr}[\rho]=1,\nonumber
\end{eqnarray}
as $\|A\|_1 = \max \text{tr}[X A]$ with a maximation
over all operators with $\| X\|_\infty=1$, according
to the variational characterisation of the trace-norm. 
In turn, obviously, 
any such $X$ with $\| X\|_\infty=1$
gives rise to the lower bound
\begin{eqnarray}
	E_{N,\text{min}}\geq  \inf && \text{tr}[X  \rho^\Gamma]-1 \\
	 \text{subject to } && \text{tr}[\rho W_i] = c_i,\nonumber\\
	&& \rho\geq 0,\nonumber\\
	&& \text{tr}[\rho]=1.\nonumber
\end{eqnarray}
We can now take an operator consisting only of 
the partial transposes of the witnesses we have 
measured,  
\begin{equation}\label{X}
	X= \sum_{i=1}^n \alpha_i   W_i^\Gamma
	+ \alpha_{n+1} \id,
\end{equation}
$\alpha_i\in\rr$ for $i=1,\dots,n+1$, 
such that $\|X \|_\infty=1$.
Then, there is nothing to minimize any more, as
$\text{tr}[\rho^\Gamma X^\Gamma]= \text{tr}[\rho X]$: 
we arrive at 
\begin{eqnarray}\label{NB}
	E_{N,\text{min}}\geq  \sum_{i=1}^n \alpha_i 
	c_i + \alpha_{n+1} -1.
\end{eqnarray}
This is indeed a very simple bound. Yet, it is a useful, 
 and tight one.
 
How can one find a suitable choice
for $\alpha_1,\dots,\alpha_{n+1}$? Any choice
such that $-\id \leq X\leq \id$ as in Eq.\ (\ref{X}) gives rise to a 
bound. In turn, one can also find the optimal choice
in an efficient manner:
The problem we encounter is,
\begin{eqnarray}
	\max &&  \sum_{i=1}^n \alpha_i 
	c_i + \alpha_{n+1} -1,
	\\
	 \text{subject to }  &&  -\id \leq X \leq \id,\nonumber\\
	&& X= \sum_{i=1}^n \alpha_i   W_i^\Gamma
	+\alpha_{n+1}\id ,\nonumber
\end{eqnarray}
as an optimization problem over 
$\{\alpha_1,\dots,\alpha_{n+1}\}$. This is an  
optimization problem we can run beforehand: it is actually
a {\it semi-definite optimization problem} \cite{Convex}, 
so an optimization problem that can 
be efficiently solved, with certifiable 
error bounds. But for the use of this criterion as 
such, one does no
longer have to solve any optimization problem.

At this point, a remark is in order concerning the 
tightness of
the constructed
bounds. Let $\rho^\Gamma$ be the partial transpose of a state on $\cc^{d_1}\otimes \cc^{d_2}$,
$N_p$ eigenvalues of which are strictly positive, 
$N_n$ eigenvalues are strictly negative, 
and $N_0$ eigenvalues take the value $0$. Then any $X$ satisfying
\begin{equation}
	\|\rho^\Gamma\|_1 = \text{tr}[X \rho^\Gamma]
\end{equation}
has a 
spectrum containing
at least $N_p$ times the value $1$ and $N_n$ times the value $-1$. In our present context, 
this means that for a given system dimension
$\cc^{d_1}\otimes \cc^{d_2}$, the above bound Eq.\ (\ref{NB}) is tight whenever the there exist
states $\rho$ such that $S_p\geq N_p$ and $S_n\geq N_n$, where $S_p$ and $S_n$ are the number of 
$\pm1$ eigenvalues of $X$, respectively. 
Then the bound is just saturated by actual physical 
states.

\begin{example}[Bound to the negativity]
{\rm 
As a very simple example, consider 
states on $\cc^2\otimes \cc^2$.
The witness we take is 
\begin{equation}\label{SWit}
	W_1 = |\phi^-\rangle\langle\phi^-|^\Gamma,
\end{equation}
which is an optimal entanglement witness, in that it is tangent
to the set of separable states. Here and in the following, $|\phi^\pm\rangle$
and $|\psi^\pm\rangle$ denotes the state vectors of the familiar 
{\it Bell states}
for two qubits. Now consider
$	X= -  2 W_1^\Gamma+ \id$, so $\alpha_1=-2$ and $\alpha_2=1$.
The matrix $X$ clearly satisfies $\|X\|_\infty=1$. Then, whenever
we get a value
	$\text{tr}[W_1\rho] = c$,
we can assert that
\begin{equation}
	E_N(\rho)\geq 2 |c|.
\end{equation}
It is also easy to see that this bound is tight: The spectrum of $X$ is
given by $\{1,1,1,-1\}$. 
A family of states saturating the bound is given by
\begin{equation}\label{SimFam}
	\rho=
	\lambda|\psi^+\rangle\langle\psi^+| +
	 (1-\lambda) |\psi^-\rangle
	\langle\psi^-|
\end{equation}
for which $E_N(\rho) = 2 |c| = |2\lambda-1|$.
For $c=-1/2$, the only state consistent with this value is
the maximally entangled state $|\psi^+\rangle\langle \psi^+|$,
yielding $E_N(|\psi^+\rangle\langle \psi^+|)=1$.
}
\end{example}

In turn, we can see what we may gain from using two
witnesses:

\begin{example}[Bound from two witnesses]
{\rm Let us take the two entanglement witnesses
	$W_i= |\phi_i\rangle\langle\phi_i|^\Gamma$, 
	$i=1,2$, where
	\begin{eqnarray}
	|\phi_1\rangle& =& 
	\frac{1}{10} |0,0\rangle
	+
	\frac{1}{10} |0,1\rangle
	+ 
	\frac{1}{5} |1,0\rangle
	+
	\left(\frac{47}{50}\right)^{1/2} |1,1\rangle,\nonumber \\
	|\phi_2\rangle& =& 
	\frac{3}{10} |0,0\rangle
	+
	\frac{1}{10} |0,1\rangle
	+ 
	\frac{1}{5} |1,0\rangle
	+
	\left(	\frac{7}{50}\right)^{1/2}
	|1,1\rangle,\nonumber\\
	\end{eqnarray}
	and $c_1=-1/3$ and $c_2=-1/6$.
	Then, we may evaluate the optimal
	bound based on each witness separately,
	and the best bound based on both simultaneously.
	From solving the semi-definite optimization problem, we 
	find in case of $W_1$,
	\begin{equation}
		E_{N,\text{min}}\geq 2/3,
	\end{equation}
	then for $W_2$, 
	\begin{equation}
		E_{N,\text{min}}\geq 1/3.
	\end{equation}
	Indeed, 
	in the combined case using $W_1 $ and $W_2$, 
	we obtain the better bound
	\begin{equation}
		E_{N,\text{min}}\geq 0.7375.
	\end{equation}
	}
\end{example}
This shows that the suitable 
processing of several witnesses at the same time can
give rise to optimized bounds. The bound arising from the
data from two witnesses is stronger than each bound resulting
from either of them.

The presented bounds are based on simple witnesses for qubit systems, but it should be
clear that the construction is general enough such that 
bounds can be identified in fact for arbitrary entanglement witnesses in any dimension.

\subsection{Convex hull measures and the conjugate function}

Many entanglement measures are defined as a {\it convex
hull} of a function, so as $\tilde f= co f$. This is
nothing but
\begin{equation}
	\tilde f(\rho)  =
	\min
	\left\{
	\sum_i p_i f(\rho_i) : \sum_i p_i \rho_i =\rho
	\right\},
\end{equation}
for states $\rho$. The most familiar example of this
sort is the {\it entanglement of formation}, for which 
this function $f$ is the reduced entropy function
\begin{equation}
	f(\rho) =  (S\circ\text{tr}_2)(\rho),
\end{equation}
where $\text{tr}_2$ is the partial trace in a bi-partite
system and $S(\rho)=-\text{tr}[\rho\log_2 \rho]$ is the 
{\it von-Neumann entropy}.
The convex hull of a function $f$
can alternatively also be written in the form
\begin{eqnarray}
	\tilde f (\rho)&=&
	\sup_X
	\bigl\{
	\text{tr}[X\rho]:
	\forall |\psi\rangle \in {\cal H}: \nonumber\\
	&& 
	\text{tr}[ |\psi\rangle\langle\psi| X]\leq 
	f( |\psi\rangle\langle\psi| )
	\bigr\}.
\end{eqnarray}
Note that we have for consistency assigned 
$f(x)=\infty$ in case of $x<0$.
Again, we aim for bounding the solution of
\begin{eqnarray}\label{Fopt}
	E_{\text{min}} = \inf &&\tilde f(\rho),\\
	 \text{subject to}  && \text{tr}[\rho W_i] = c_i,\nonumber\\
	&& \rho\geq 0,\nonumber\\
	&& \text{tr}[\rho]=1,\nonumber
\end{eqnarray}
$i=1,\dots,n$. 
We can make use of the {\it conjugate function}
[42--44], also known as the 
{\it Legendre transform}: This 
is defined as
\begin{eqnarray}\label{NC}
	 f^\ast (X) &=& \sup_{\rho\in {\cal S}({\cal H})}
	\bigl\{
	\text{tr}[\rho X] - f(\rho)
	\bigr\},
\end{eqnarray}
again ${\cal S}({\cal H})$ denoting state space,
which is
\begin{eqnarray}\label{conjp}
	f^\ast (X)&=&\sup_{|\psi\rangle\in{\cal H} }
	\bigl\{
	\text{tr}[  |\psi\rangle\langle\psi| X] - 
	f(|\psi\rangle\langle\psi|)
	\bigr\} 
\end{eqnarray}
for concave functions $f$.
In turn, the conjugate function of the conjugate is 
the convex hull of the function itself \cite{Convex}: 
In other words,
since the entanglement of formation is the convex hull of
the reduced entropy function itself, we have that
\begin{equation}
	f^{\ast\ast}(\rho) =  \tilde f(
	\rho),
\end{equation}
where
\begin{equation}
	f^{\ast\ast}(\rho) = \sup_X
	\bigl\{
	\text{tr}[  \rho X] - 
	f^\ast (X)
	\bigr\} .
\end{equation}
By definition, $f^\ast (X+\alpha \id) = 
f^\ast (X) +\alpha $ for any $X$. Now, for any 
\begin{equation}
	X= \sum_{i=1}^n \alpha_i W_i ,
\end{equation}
we indeed arrive at the bound
\begin{equation}
	E_{\text{min}}\geq \sum_{i=1}^n c_i \alpha_i  
	- f^\ast (X),
\end{equation}
in terms of the conjugate function $f^\ast$ of $f$. In this
way, we do not have to evaluate the convex hull 
explicitly. 

Moreover, the bounds constructed in this way are always tight. It follows 
from the duality of the convex hull of the function 
and its Legendre transform that the bounds are tight when
varying over all $\{\alpha_1,\dots, \alpha_n\}$. There always exists a
state $\rho$ satisfying $E_{\text{min}}= f^{\ast\ast}(\rho)$, so 
for example $E_{F,\text{min}}= E_F(\rho)$ for the entanglement of 
formation. In this sense, the given bounds are the best possible bounds of this 
form.

In case a symmetry can be identified, the estimation of 
the conjugate function of a given function can be simplified.
To bring the conjugate function into a form that is more accessible to 
numerical assessments, we can proceed as follows:
If $f=(g\circ {\text{tr}}_2)$, and $g$ is concave,
we can define
\begin{eqnarray}
	g(\rho) &=& \inf_Y \{ \text{tr}[Y\rho] - g'(Y)\},\\
	g'(Y) &=& \inf_\rho \{ \text{tr}[Y\rho] - g(\rho)\},
\end{eqnarray}
and can write the above conjugate function as
(assuming that $X$ and $Y_1$ are Hermitian)
\begin{eqnarray}
    f^\ast (X) &=& \sup_\rho \sup_{ Y_1 } 
    \left\{ \text{tr}[(X - (Y_1 \otimes \id))\rho]+ g'(Y_1) \right\} \nonumber
    \\
    &=& \sup_{ Y_1 } \left\{ \lambda_{\max}(X - (Y_1 \otimes \id)) + g'(Y_1) \right\}.
\end{eqnarray}
For the entropy function $g(x)= -x\log_2 x$, for example,
the conjugate $g'$ is known, and one finds \cite{Convex}
\begin{equation}
	g'(Y_1)= -\log_2
	\text{tr}[\exp({-Y_1\log 2})].
\end{equation} 
In this form, the problem is in a suitable form for such numerical assessments.
The resulting bound is then a combination of
the numerically evaluated expression and the value for $c$ from the actual data.
In practice, this numerical evaluation amounts to a global optimization problem,
which can, for a small number of parameters in typical problems in the quantum 
information context, be solved for an arbitrary witness. 
Also, semi-definite relaxations
as in Refs.\ \cite{Hierarchy,Lasserre} 
readily give rise to certifiable bounds. 
  
As an example, let us look at the {\it entanglement of formation},
and a single witness $W_1$. Then, 
\begin{eqnarray}
	E_{F,\text{min}}	&\geq &
	\alpha_1 c - 
	f^\ast(\alpha_1  W_1 ),
\end{eqnarray}
with $\alpha_1 \in \rr$, so any choice for 
$\alpha_1$ delivers a bound. Obviously, 
an optimal bound is achieved using
\begin{eqnarray}
	E_{F,\text{min}} \geq \text{sup}_{\alpha_1 }
	\left\{
	\alpha_1 
	c  - 
	f^\ast(\alpha_1  W_1 )\right\}
	.
\end{eqnarray}
Similarly, more than a single 
witness can be considered. So one needs to 
find good upper bounds to $f^\ast(\alpha_1  W_1 )$.

\begin{example}[Bound to the entanglement of formation]
{\rm
This becomes particularly simple for 
witnesses of the form
	$W_1=|\phi\rangle\langle\phi|^\Gamma$
	in $\cc^d\otimes \cc^d$,
	for entangled state vectors $|\phi\rangle$.
	We consider the conjugate function $f^*$,
	evaluated at $\alpha_1 W_1$. 
	It is easily seen from Eq.\ (23)
  that, for any entanglement measure, the conjugate function 
  is invariant under local unitaries. 
  Then, without loss of generality 
	$|\phi\rangle$ can be taken to be of Schmidt form
	\begin{equation}\label{Schmidt}
	|\phi\rangle=\sum_{i=1}^d
	\xi_i |i,i\rangle.
	\end{equation}
	The partial transpose $|\phi\rangle\langle\phi|^\Gamma$
	gives rise to the form
	\begin{equation}
		|\phi\rangle\langle\phi|^\Gamma= \sum_{j,k=1}^d
		\xi_j \xi_k |j,k\rangle\langle k,j| ,
	\end{equation}
	so in a product basis a direct sum of $1\times 1$
	and $2\times 2$ matrices.
	We seek the maximal value of
	$\alpha_1 \text{tr}[W_1|\psi\rangle\langle\psi| ] $
	and a minimal value for $f(|\psi\rangle\langle\psi|)$.
	Let
	\begin{equation}
	w= \max_{i,j\in\{1,\dots,d\},\,i<j} \{\xi_i
	\xi_j\},
	\end{equation}
	so $w= \xi_k\xi_l$ for some $k,l=1,\dots,d$.
	It is not difficult to see that then the optimal 
	state vector $|\psi\rangle$ takes the form 
	\begin{equation}
		|\psi\rangle = 
		a |k,l\rangle - (1-a^2)^{1/2} |l,k\rangle
	\end{equation}
	for some $a\in[0,1]$. This state vector gives
	rise to
\begin{equation}
	\text{tr}[- W_1|\psi\rangle\langle\psi| ]=
	 2  a (1-a^2)^{1/2} w
\end{equation}
and
$f(|\psi\rangle\langle\psi|)=
	-a^2 \log_2 a^2 - (1-a^2) \log_2 (1-a^2)$,
defining the concave 
(classical entropy) function $h(x)=-x\log_2 x-(1-x)\log_2(1-x)$
and the concave function $g(x)=({x(1-x)})^{1/2}$,
\begin{eqnarray}
    f^\ast(\alpha_1 W_1)
    &=& \sup_{a\in [0,1]}\biggl\{ - 2 \alpha_1 w a (1-a^2)^{1/2} 
    \nonumber\\
    &+&  
    a^2 \log_2 a^2 + (1-a^2) \log_2 (1-a^2) \biggr\}\nonumber \\
    &=& \sup_{p\in [0,1]}\bigl\{  - 2 \alpha_1 w g(p) -h(p)\bigr\}.
\end{eqnarray}
We can distinguish three regimes.
Define the parameter $b=-2\alpha_1 w$ and the function $z(p)=b g(p)-h(p)$.
The second derivative of $z$ is given by
\begin{equation}
z''(p) = \frac{1}{\log(2)p(1-p)} - \frac{b}{4(p(1-p))^{3/2}}.
\end{equation}
The function $z$ is convex iff $z''$ is non-negative for all 
$p\in[0,1]$, which occurs
when 
\begin{equation}
	b\le\min_p \frac{4({p(1-p)})^{1/2} }{\log(2)} = 0.
\end{equation}	
The function is concave when $z''$ is non-positive for all $p$, 
which occurs when
\begin{equation}
	b\ge\max_p {4 {p(1-p)}^{1/2}   }/{\log(2)} = 2/\log(2) =2.88539.
\end{equation}
In between these values, $z$ is neither convex nor concave.
If $z$ is convex ($b\le0$), its supremum occurs at one of the extreme points,
either $p=0$ or $p=1$. But of course, either one 
gives the same value, namely $0$.
If $z$ is concave ($b\ge 2/\log(2)$), 
it has one supremum. By the even symmetry of $z$ around
$p=1/2$, the supremum must occur at $p=1/2$, yielding as supremum value $(b-2)/2$.
For determining the supremum in the case $0\le b\le 2/\log(2)$, a transcendental equation has to be solved.
The supremum as function of $b$ can be approximated from above by the polynomial
\begin{eqnarray}
0.001876 b &+&
    0.008239 b^2 + 0.019733 b^3 \\ &-&
    0.005649 b^4 + 0.001430 b^5. \nonumber
\end{eqnarray}
The average error of this approximation is $0.00017$.
We may take $\alpha_1= -1/w$, then 
$f^\ast(\alpha_1 W_1) \leq c_0$ with 
$c_0=0.14985$. Therefore, we 
obtain the bound
\begin{equation}
	E_{F,\text{min}}\geq
	|c|/|w| - c_0.
\end{equation}
}
\end{example}
To emphasize that again, we do not 
assume the ``true state'' to be detected to be
known or accessible. For completeness, we do 
elaborate on an example showing the tightness 
of the bound.  For example, for the  
family of states in Eq.\ (\ref{SimFam}) for $\lambda\in[1/2,1]$,
and for the witness
$W_1 = |\phi^-\rangle\langle\phi^-|^\Gamma$ as
in Eq.\ (\ref{SWit}),
we find $2|c|=|2\lambda -1|$, $w=1/2$, 
and in fact 
\begin{equation}
	E_F(\rho) = h\left(1/2 + (\lambda (1-\lambda))^{1/2} \right),
\end{equation}
which has to be compared with
\begin{equation}
	E_{F,\text{min}}\geq
	|2\lambda-1 | - c_0.
\end{equation}
As can easily be seen, this is a very good lower
bound (as a tangent the best possible affine bound),
and the bound is tight for $\lambda = 0.7056$.

\begin{example}[Second 
bound to the entanglement of
formation] 
{\rm 
Let us consider  a witness of the common form
\begin{equation}
	W_1 = a \id - b|\phi\rangle\langle\phi|
\end{equation}
with $a,b>0$. The conjugate function can be easily written as
\begin{eqnarray} \label{f*1}
	f^\ast (\alpha_1 W_1)=\sup_{|\psi\rangle\in{\cal H} }
	\bigl\{ \alpha_1 a - \alpha_1 b |\braket{\phi}{\psi}|^{2} - 
	f(\ket{\psi}\bra{\psi}) \bigr\}.
\end{eqnarray}
We assume $\ket{\phi}$ to be in 
its Schmidt form, given by Eq.\ 
(\ref{Schmidt}). The state vector 
$\ket{\psi}$ might also be written as
\begin{equation}
\ket{\psi} = \sum_{i=1}^{d}\mu_{i} \ket{i', i'},
\end{equation}
where the basis $\{ \ket{i'} \}$ is not
necessarily equal to the Schmidt basis $\{ \ket{i} \}$ of 
$\ket{\phi}$. One can thus minimize 
over the basis $\{ \ket{i'} \}$ and the Schmidt coefficients 
$\mu_{i}$. The last term in the right hand side of Eq.\ 
(\ref{f*1}) clearly does not depend on $\{ \ket{i'} \}$. 
In turn, given a fixed set of Schmidt coefficients $\mu_i$, 
this implies that the optimal basis will be the one which 
maximizes the overlap $|\braket{\phi}{\psi}|^{2}$. 
It can be easily shown that the maximum is obtained 
when choosing $\{ \ket{i'} \}$ to be equal to the Schmidt 
basis of $\ket{\phi}$. Therefore, we are left with an 
easier maximization problem, over the Schmidt 
coefficients only, given by
\begin{eqnarray} \label{maxSch}
	f^\ast (\alpha_1 W_1)&=&\sup_{ \{ \mu_i \} } 
	\biggl\{ 
	\alpha_1 a - \alpha_1 b \biggl(\sum_{i}\xi_i \mu_i \biggr)^{2} \\ 
	 &+& \sum_{i}\mu_i^2\log_2(\mu_i^2) \biggr\}. \nonumber
\end{eqnarray}
Although it is not possible to solve Eq.\ (\ref{maxSch}) 
analytically in terms of $\alpha_1, a$ and $b$ for all choices, 
it can be easily numerically evaluated. For example, let us 
consider $a = \alpha_1 = 1$, $b = 3/2$, and
\begin{equation}
\ket{\phi} = \left({\frac{1}{3}}\right)^{1/2}
\ket{0, 0} + \left({\frac{2}{3}}\right)^{1/2} 
\ket{1, 1},
\end{equation}
so $\xi_1=(1/3)^{1/2}$ and $\xi_2 = (2/3)^{1/2}$.
We then
get the following bound for the entanglement of 
formation
\begin{equation}
	E_{F,\text{min}}\geq
	|c| - 0.5550.
\end{equation}
}
\end{example}

\begin{example}[Bounds from Renyi entropies]
{\rm 
Since for the entropy function $f$, we have that 
\begin{equation}
	f(x) \geq g_q(x) = \frac{1}{1-q}\log_2(x^q)
\end{equation}
for $q>1$, we get an upper bound to 
$f^\ast (X)$ as
\begin{equation}
	f^\ast (X) \leq \sup_{|\psi\rangle}
	\bigl\{
	\text{tr}[  |\psi\rangle\langle\psi| X] - 
	g_q(|\psi\rangle\langle\psi|)
	\bigr\} .
\end{equation}
The function $g_q$ is no longer concave, but
we nevertheless get an appropriate bound when
optimizing over pure states. Particularly useful is the
case of $q=2$, when we merely need to evaluate
$\text{tr}[\text{tr}_2[|\psi\rangle\langle\psi|]^2]$.
}
\end{example}

Further bounds can, e.g., 
be found in Refs.\ \cite{Breuer,Chen}.

\subsection{Remarks on exploiting symmetry}

If one has a witness which is invariant under
a local symmetry group, one can in instances 
simplify the evaluation of good
bounds under the constraint provided by the entanglement
witness: One can take the Haar average with respect
to that group, which will always diminish the degree of
entanglement. So a twirling with respect to a, for example, 
$U\otimes U$,
$U\otimes U^\ast$,  or $O\otimes O$-symmetry, or
one corresponding to $SU(2)$ or symmetric group
representations, can only
give a lower bound \cite{Vollbrecht}: For any
convex entanglement monotone $f$, 
\begin{eqnarray}
        E_{\text{min}} = \inf &&
        f(\rho)\\
	\text{subject to} &&
	\text{tr}[\rho W_1]=c_1,\nonumber\\
	&& \rho \text{ is a symmetric state}.\nonumber
\end{eqnarray}
Hence, we have to evaluate an
entanglement measure under symmetry 
[49--53],
given the constraint. 

\begin{example}[Symmetry] {\rm 
To give a very simple example, let us consider a
witness of the form 
\begin{equation}
	W_1 = a\id + b |\phi^-\rangle\langle\phi^-|^\Gamma
\end{equation}	
for some $a,b\in \rr$ and for states on $\cc^2\otimes \cc^2$.
Since $|\phi^-\rangle\langle\phi^-|$ is a $U\otimes U$-symmetric 
state, this witness is $U\otimes U^\ast$-symmetric. Therefore, we can optimize
the bound with respect to $U\otimes U^\ast$-symmetric states, which is the 
one-dimensional convex set
\begin{equation}
	\rho = \lambda \id/4+ (1-\lambda)|\psi^+\rangle\langle\psi^+|,
\end{equation}
for $\lambda\in[0,1]$. The entanglement of formation $E_F(\rho)$ of such symmetric states 
$\rho$, in turn, is known \cite{Vollbrecht}. 
}
\end{example}

\subsection{Concurrence}

An interesting example where the conjugate function can be 
analytically 
calculated is the concurrence of two qubits \cite{Wootters}. 
Let us define the following basis for $\cc^2 \otimes \cc^2$,
\begin{eqnarray}
\ket{\Psi_0} = \frac{1}{\sqrt{2}}(\ket{0,0} + \ket{1,1}), 
\ket{\Psi_1} = \frac{i}{\sqrt{2}}(\ket{0,0} - \ket{1,1}), \\
\ket{\Psi_2} = \frac{i}{\sqrt{2}}(\ket{0,1} + \ket{1,0}), 
\ket{\Psi_3} = \frac{1}{\sqrt{2}}(\ket{0,1} - \ket{1,0}).
\end{eqnarray}
As is well-known, a 
general two-qubit pure state can then be written as 
\begin{equation} \label{exp}
	\ket{\phi(c)} = \sum_{i=0}^{3}c_i \ket{\Psi_i}.
\end{equation}
The {\it concurrence} of a pure state is defined as
\begin{equation}\label{con} 
C(|\psi\rangle\langle \psi|) 
= \left| \sum_{i=0}^{3}c_i^2 \right|,
\end{equation}
and extended to mixed states by a convex hull construction. 
The importance of the concurrence is twofold. 
On one hand, it is intimately related to the entanglement 
of formation of two qubits. Indeed, given the concurrence 
of a two qubit state $\rho$, its entanglement of formation 
reads \cite{Wootters}
\begin{equation} 
	E_{F}(\rho) = H\left(\frac{1}{2}\left(1 + (1 - C(\rho)^2)^{1/2}
	\right)\right).
\end{equation}
On the other hand, an analytical expression for the concurrence of a general two qubits mixed state is known \cite{Wootters}, which in turn implies an analytical formula for the entanglement of formation. 

The first interesting  bound to the concurrence based on a witness, 
derived in  Ref.\ \cite{VerstraetePhD}, is given by
\begin{equation} \label{verstraete}
	C(\rho) = 
	\max \left
	\{ 0, -\min_{A \in SL(2,\cc)}\text{tr}
	[\ket{A}\bra{A}^{\Gamma}\rho]
	\right \},
\end{equation}
where $\ket{A}$ denotes the unnormalized state 
vector $(A \otimes \id)
\ket{I}$ with $\ket{I} = \ket{0, 0} + \ket{1, 1}$ 
and $A$ is any (in its determinant normalized)
$2 \times 2$ invertible matrix. 
It is thus seen that any witness $W_1$ of the form
\begin{equation} \label{optwit}
	W_1= \ket{A}\bra{A}^{\Gamma}
\end{equation}
provides a lower bound to the concurrence. 
Note that this class of witnesses is exactly the 
class of optimal entanglement witnesses of two qubits, 
where optimality refers to the robustness with respect 
to white noise (mixing with the identity). 

Although 
Eq.\ (\ref{verstraete}) constitutes a
 useful tool to estimate $C$, it has the drawback 
 that only witnesses of the restricted form given by Eq.\ 
 (\ref{optwit}) can be used. On the other hand, the 
 method based on conjugate functions, oulined in the 
 previous subsection, can be applied to any entanglement 
 witness. We now aim at showing that 
 the conjugate function of the concurrence can be evaluated analytically.

The concurrence can be expressed as the convex-hull of the function 
\begin{equation}
	f(\rho) = \left(2(1 - \text{tr}[\text{tr}_2[\rho]^{2}])\right)^{1/2},
\end{equation}
defined on states $\rho$.
It is easy to check that $f$ is concave.
This in turn implies that the supremum in 
Eq.\ (\ref{conjp}) 
can be calculated over pure states only.
The conjugate function
$f^{*}$ can be expressed as the following 
optimization problem,
\begin{eqnarray}
	f^{*}(X) &=& \sup_{\{ c_i \}} \sum_{i, j}\bra{\Psi_i}X\ket{\Psi_j}c_i^* c_j - |\sum_{c_i} c_i^2|,\\
	\text{subject to}&& \ket{\phi} = \sum_i c_i \ket{\Psi_i}
	\text{ is normalized.}\nonumber
\end{eqnarray}
The optimal solution, as a function of $X$, although this
not being a convex problem, can be readily 
evaluated with the help of a computer algebra 
program \cite{Alg}.  

\subsection{Further convex roof measures}

Note that we considered the entanglement of formation as
an example for a ``convex roof measure''. There are other
important measures of entanglement in the multi-partite
context, where the presented ideas can be applied.
This applies in particular to the
{\it geometric measure of entanglement}. This is a
measure for entanglement in the multi-partite case, 
which is defined
for pure states as \cite{Geometric}
\begin{equation} \label{geo}
	E_G(|\psi\rangle\langle\psi|) =
	\inf_\rho \| |\psi\rangle\langle\psi| - \rho\|_2,
\end{equation}
where $\|A\|_2 = \text{tr}[A^2]$ is the Hilbert-Schmidt
norm, and the infimum is taken with respect to 
all pure product states. The extension to mixed states
is done via a convex roof construction. Similarly,
the global entanglement of Ref.\ \cite{Global}
may be considered. Both quantities are proper
multi-partite entanglement monotones. For a survey
on multi-partite entanglement measures, see, e.g.,
Refs.\ \cite{Multi3,PlenioMeasures}.

\subsection{Robustness}

Given a bi-partite 
or multi-partite state $\rho$, its {\it generalized robustness of entanglement} --  
introduced in Refs.\ \cite{VidalTarrach, Steiner}
-- is defined as the minimal   $s>0$ such that 
the state
\begin{equation} \label{primal}
	\omega= \frac{\rho + s \sigma}{1 + s}
\end{equation} 
is separable, where $\sigma$ is another arbitrary state. This measure can be interpreted as the minimum amount of noise necessary to wash out completely the quantum correlations initially present 
in the state $\rho$. In addition, the generalized robustness also has the operational interpretation for bi-partite systems as the usefulness of the state in question as an ancilla in teleportation protocols 
\cite{Brandao2} and is a multi-partite entanglement monotone \cite{Brandao1}.  

For our purposes,
a very convenient representation of the 
generalized robustness, obtained as 
the \textit{Lagrange dual} form 
\cite{Convex}
of Eq.\ (\ref{primal}) \cite{Brandao1}, is 
\begin{equation}
	E_R(\rho) = 
	\max \left\{ 0, - \min_{W} \text{tr}[W \rho] \right\}, 
\end{equation}
where $W$ is
varied over the set of witnesses with maximum eigenvalue smaller than unity ($W \leq \id$). Note that by considering different sets of witnesses, one can quantify all the different kinds of multi-partite entanglement [27--29].

As discussed in Refs.\ \cite{Brandao1, Cavalcanti}, it follows directly from 
Eq.\ (\ref{dual}) that the expectation value of any measured witness 
$W$ gives rise to a useful lower bound to the generalized robustness.  
Then, when $\text{tr}[W\rho]=c<0$,
\begin{equation} \label{dual}
	E_{R,\text{min}}(\rho) \geq  |c|/
	{\lambda_{\max}(W)}.
\end{equation}
In full generality, this same approach can be applied to any entanglement measure which can be expressed as
\begin{equation} \label{witnessed}
	E(\rho) = 
	\max \left\{
	0, -\min_{W \in {\cal M}} \text{tr}[W\rho] \right\},
\end{equation}
where ${\cal M}$ is the intersection of the sets of entanglement witnesses with some other set (e.g., the set 
$W \leq \id$). Interestingly, 
several other well-known entanglement quantifiers, 
such as the \textit{best separable approximation} 
\cite{Karnas}, the \textit{Rains fidelity of teleportation} 
\cite{Rains}, and the {\it concurrence} (see Eq.\ 
(\ref{con})), 
fit into this classification. Here for concreteness 
we focus on the {\it generalized robustness} and on 
the \textit{random robustness}, which 
we discuss in the sequel.     

A source of noise often 
considered in experiments is 
so-called white noise, in which the 
initial state $\rho$ is driven 
to a state of the form
\begin{equation}
	\rho\mapsto \rho + s \frac{\id}{D},
\end{equation}
where $s$ is related to the amount of noise introduced in the system. Here $D$ stands for the dimension of the Hilbert space which $\rho$ acts on. In this sense, it is interesting to ask what is the maximal tolerance of an entangled state to white noise, before all its initially entanglement is transformed into merely classical correlations. The {\it random robustness}
\cite{VidalTarrach} is exactly such a 
quantity.  In the framework of Eq.\ 
(\ref{witnessed}), we can express it 
as the minimization over the set of entanglement 
witnesses with trace equal to $D$. Hence, every entanglement 
witness $W$ can be used to lower bound it as
\begin{equation}
	E_{r,\text{min}}(\rho) \geq  \frac{D|c|}{\text{tr}[W]},
\end{equation}
again with $\text{tr}[W\rho]=c$.

\begin{example}[Tri- and quadripartite photonic
entanglement]
{\rm 
As an example, we consider two multi-partite witnesses which have been measured in the photonic parametric-down-conversion experiment of Ref.\ \cite{WeinDetect}, so $N=3$ and $N=4$. 
Consider the following multi-partite pure states vectors
\begin{equation}
\ket{W} = \frac{1}{\sqrt{3}} (\ket{0,0,1} + \ket{0,1,0} + \ket{1,0,0})
\end{equation}
and
\begin{eqnarray}
&&\ket{\Psi^{(4)}} = \frac{1}{\sqrt{3}} ( \ket{0,0,1,1} + \ket{1,1,0,0}  \\ 
                  &-& \frac{1}{2}(\ket{0,1,1,0} + \ket{1,0,0,1} + \ket{0,1,0,1} + 	\ket{1,0,1,0})).
                  \nonumber
\end{eqnarray}
Then the two associated multi-partite entanglement witnesses 
which have been 
measured are given by
\begin{eqnarray}
	W_{W} &=& \frac{2}{3}\id - \ket{W}\bra{W},\\
	W_{\Psi^{(4)}} & =&
	 \frac{3}{4}\id - \ket{\Psi^{(4)}}\bra{\Psi^{(4)}}.
\end{eqnarray}
Whereas the witness $W_{W}$ detects 
genuine tri-partite entanglement, having positive values on separable and bi-partite entangled states, the operator $W_{\Psi^{(4)}}$ witnesses 
genuine four-partite entanglement, being positive on 
separable, bi-separable, and tri-separable states.
The measured expectation values in turn are \cite{WeinDetect},
\begin{eqnarray}
	\text{tr}[W_{W}\rho] &=& - 0.197 \pm 0.018,\\
	\text{tr}[W_{\Psi^{(4)}}\rho] &=& - 0.151 \pm 0.010. 
\end{eqnarray}
Hence we readily have the following estimates on the robustness of the, a priori unknown, measured states $\rho_1$ and $\rho_2$, consiting of three and four parties respectively:
\begin{eqnarray}
	E_{R,\text{min}}(\rho_1) &\geq & 
	 0.2955 \pm 0.027, \hspace{0.2 cm} E_{r,\text{min}}(\rho_1) \geq 0.360 \pm 0.096,\nonumber\\
	 E_{R,\text{min}}(\rho_2) &\geq&  
	 0.201 \pm 0.013, \hspace{0.2 cm} E_{r,\text{min}}(\rho_2) \geq  0.220 \pm 0.021.\nonumber\\
\end{eqnarray}
}
\end{example}

\begin{example}[Four-photon graph state] 
{\rm 
In Ref.\
\cite{NewWein}, entanglement witnesses have been
employed to characterize optical four-photon
graph states[70--74]
that have been prepared from 
entangled photon pairs, followed by a controlled-phase
gate (compare also Ref.\ \cite{ZeilingerCluster}). 
For the four-photon cluster state 
\cite{BriegelPersistentEntanglement},
$N=4$, the given witness is
\begin{eqnarray}
	W_{{\cal C}^{(4)}} &=& 3 \id - \frac{1}{2}
	\left(
	Z^{(1)} Z^{(2)} + \id
	\right)
	\left(
	Z^{(2)} X^{(3)}  X^{(4)} + \id
	\right) \nonumber\\
	&-&
	 \frac{1}{2}
	\left(
	X^{(1)} X^{(2)} Z^{(3)} + \id
	\right)
	\left(
	Z^{(3)} Z^{(4)}    + \id
	\right) .
\end{eqnarray}
The maximal theoretical value is 
$\text{tr}[W_{{\cal C}^{(4)}}\rho]=-1$, the 
measured value is
\begin{equation}
\text{tr}[W_{{\cal C}^{(4)}}\rho]=-0.299\pm 0.050.
\end{equation}
This gives rise to
\begin{eqnarray}
	E_{R,\text{min}}&\geq&  
	0.0997\pm 0.0167  , \\
	E_{r,\text{min}}&\geq& 0.1120\pm 0.020.
\end{eqnarray}
}
\end{example}

\begin{example}[Quantum byte] 
{\rm 
In the recent
spectacular experiment of Ref.\  \cite{Byte},
$8$ ions have been prepared in a multi-particle
entangled state. The multi-particle entanglement
has in turn been demonstrated using the concept
of entanglement witnesses. 
In order to introduce the multi-partite entanglement witnesses that have been
measured, we have to consider the $N$-partite $W$ states
\begin{eqnarray}  
\ket{W_N} &=& (\ket{0, \dots, 0, 0, 1} + \ket{0, \dots, 0, 1, 0} \\ &+& \ket{0, \dots, 1, 0, 0} + \dots + \ket{1, \dots, 0, 0, 0})/\sqrt{N}.\nonumber
\end{eqnarray}
Define the $N$-qubit state vectors 
$\ket{BS_i} = \ket{D_i}\otimes \ket{W_{N-1}}$, 
which consist of $\ket{0}$ on the $i$-th 
qubit and the state vector $\ket{W_{N-1}}$ 
on the remaining qubits, and the 
corresponding operators
\begin{equation}
{\cal Q}_N = 10 \ket{W_N}\bra{W_N} - \beta_N \sum_{i=1}^N \ket{BS_i}\bra{BS_i},
\end{equation}
where $\beta_N$ is a fixed real number for each value. 
Next, define $\gamma_N = \max_{\ket{\Psi} = \ket{a}\otimes \ket{b}}\bra{\Psi}{\cal Q}\ket{\Psi}$, where 
$\ket{\Psi}$ ranges over all possible 
bi-separable state vectors 
\cite{Multi1,Multi3} 
with respect to all possible bi-partitions. 
The witnesses are then given by
\begin{equation}
	W_{\text{Byte}} = \gamma_N \id - {\cal Q}_N.
\end{equation}
They hence classify tri-partite entanglement. As explained in 
Ref.\ \cite{Byte}, the expectation values reported refer to the 
normalized versions of ${W}_{\text{Byte}}$ with 
$\text{tr}[{W}_{\text{Byte}}] = 2^N$, 
where $N$ is the number of parties of the state. 
Therefore, we can readily read them as lower 
bounds to the random robustness of the 
unknown $3$, $4$, $5$, $6$, $7$, and $8$ qubit states:
\begin{eqnarray}
E_{r, \text{min}}(3) &\geq&  0.532, \hspace{0.2 cm} 
E_{r, \text{min}}(4) \geq 0.460, \hspace{0.2 cm}  \\
E_{r, \text{min}}(5) &\geq& 0.202, \hspace{0.2 cm}    
E_{r, \text{min}}(6) \geq  0.271, \hspace{0.2 cm}  \\
E_{r, \text{min}}(7) &\geq& 0.071, \hspace{0.2 cm}  
E_{r, \text{min}}(8) \geq 0.029. \hspace{0.2 cm}  
\end{eqnarray} 
}
\end{example}

Needless to say, the same discussion can also be 
performed based on other multi-partite measures of
entanglement, such as the geometric measure of entanglement.

\section{Conclusions}

In this work, we have introduced quantitative 
bounds to entanglement measures, on the basis of
expectation values of 
entanglement witnesses. In this way, 
quantities that are frequently measured in order to detect
entanglement in experimental settings can be
augmented with a stronger, quantitative statement
on the degree of entanglement. 
In most instances, this
does not require any additional effort at all, but these
quantitative bounds may even 
be added in retrodiction. Several measures
of entanglement have been considered.
Needless to say, similar methods can also be made use of if
one has additional knowledge at hand about
the system, say, from correlation measurements.

We have considered the concept of the conjugate function
in this context, and have presented a number of new bounds 
to entanglement measures. For the negativity, simple and 
very useful bounds emerged from a variational principle.
We have discussed several examples taken from 
experimental settings, both from the context of linear optics,
as well as of trapped ions. In this way, we have sharpened
the notion that if a test for entanglement is indeed 
violated to a large extent, then the degree of entanglement
can be expected to be large.

\section{Acknowledgements}

We would like to thank O.\ G{\"u}hne, M.\ Reimpell, and
R.F.\ Werner for valuable
discussions on the subject of the paper and P.\ Hyllus
for very helpful comments on the manuscript.
This work has been coordinated in submission 
with their independent work Ref.\ \cite{BS}.
Note also that the  independent work Ref.\
\cite{Martin} (which was made available slightly
later on the preprint server)
is similar in its spirit, and strongly and nicely 
complements the present work in 
that the role of correlation measurements is 
emphasized and studied in great detail.
The semi-definite program was 
programmed using the packages {\it SeDuMi} and {\it Yalmip}.
This work has been supported by the DFG
(SPP 1116, SPP 1078), the EU (QAP), 
the EPSRC, the QIP-IRC, 
Microsoft Research, the Brazilian agency 
Conselho Nacional de Desenvolvimento 
Cient\'ifico e Tecnol\'ogico (CNPq), and
the EURYI Award Scheme.


\begin{thebibliography}{99}

\bibitem{Horo}
	M.\ Horodecki, P.\ Horodecki, and R.\ Horodecki,
	Phys.\ Lett.\ A {\bf 223}, 1 (1996).
	
\bibitem{Terhal}
	B.M.\ Terhal, Phys.\ Lett.\ A {271}, 319 (2000).

\bibitem{TerhalSurvey}	
	B.M.\ Terhal, 
	J.\ Th.\ Comp.\ Sc.\ {\bf 287}, 313 (2002).

\bibitem{Reflect}
	D.\ Bruss, J.I.\ Cirac, P.\ Horodecki, 
	F.\ Hulpke, B.\ Kraus, M.\ Lewenstein, 
	and A.\ Sanpera,
	J.\ Mod.\ Opt.\ {\bf 49}, 1399  (2002).
	
\bibitem{WeinDetect}
        M.\ Bourennane, M.\ Eibl, C.\ Kurtsiefer
        S.\ Gaertner,
        H.\
        Weinfurter,
        O.\ G{\"u}hne, P.\ Hyllus,
        D.\ Bruss, M.\ Lewenstein, and A.\ Sanpera,
        Phys.\ Rev.\ Lett.\ {\bf 92}, 087902 (2004).
        
\bibitem{NewWein}
	N.\ Kiesel, C.\ Schmid, U.\ Weber, O.\ G{\"u}hne, 
	G.\ Toth, R.\ Ursin, and H.\ Weinfurter,
	Phys.\ Rev.\ Lett.\ {\bf 95}, 210502 (2005).
 
 
\bibitem{Byte}
	H.\ H{\"a}ffner, W.\ H{\"a}nsel, C.F.\ Roos, J.\ Benhelm,
	D.\ Chek-al-kar, M.\ Chwalla, T.\ K{\"o}rber, U.D.\ Rapol,
	M.\ Riebe, P.O.\ Schmidt, C.\ Becher, O.\ G{\"u}hne,
	W.\ D{\"u}r, and
	R.\ Blatt, Nature {\bf 438},  643 (2005).

\bibitem{Kaler}
	C.\ Becher, J.\ Benhelm, D.\ Chek-Al-Kar, M.\ Chwalla, H.\ H{\"a}ffner, W.\ H{\"a}nsel, 
	T.\ K{\"o}rber, A.\ Kreuter, G.P.T.\ Lancaster, T.\ Monz, 
	E.S.\ Phillips, U.D.\ Rapol, M.\ Riebe, C.F.\ Roos, C.\ Russo, F.\ Schmidt-Kaler, 
	and R.\ Blatt, {\it Entanglement of trapped ions},
	Proc.\ 17th  Int.\  Conf.\  Laser Spectroscopy, 
	Cairngorms National Park, Scotland, 2005, Eds. 
	E.A.\ Hinds, A.\ Ferguson, and E.\ Ries, World Scientific.
	
\bibitem{Toth}
	G.\ Toth and O.\ G{\"u}hne, 
	Phys.\ Rev.\ Lett.\ {\bf 94}, 060501 (2005).

\bibitem{Guehne}
	O.\ G{\"u}hne, P.\ Hyllus, D.\ Bruss, A.\ Ekert, 
	M.\ Lewenstein, C.\ Macchiavello, and A.\ Sanpera,
	J.\ Mod.\ Phys.\ {\bf 50}, 1079 (2003).
	
 
\bibitem{Hierarchy}
         J.\ Eisert, P.\ Hyllus, O.\ G{\"u}hne, and M.\ Curty,
        Phys.\ Rev.\ A {\bf 70}, 062317 (2004).
         
\bibitem{Doh}
         A.C.\ Doherty, P.A.\ Parrilo, and F.M.\ Spedalieri,
        Phys.\ Rev.\ A {\bf  71}, 032333 (2005).
	
\bibitem{Gab}
 	G.A.\ Durkin and C.\ Simon,
	Phys.\ Rev.\ Lett.\ {\bf 95} 180402 (2005).
	
\bibitem{NoteThat}
	Note that in the independent work
	Ref.\ \cite{BS}, O.\ G{\"u}hne,
	M.\ Reimpell, and R.F.\ Werner came to similar
	conclusions, and the two submissions have
	been coordinated.
	
\bibitem{BS}
	O.\ G{\"u}hne, M.\ Reimpell, and R.F.\ Werner, 
	quant-ph/0607163.


   		
\bibitem{OW}
	 M.\ Lewenstein, B.\ Kraus, J.I.\ Cirac, and P.\ Horodecki,
	 Phys.\ Rev.\ A {\bf 62}, 052310 (2000).
   
 \bibitem{Gurvits}
        L.\ Gurvits, {\it Proceedings of the thirty-fifth ACM 
        symposium on theory of computing},
        San Diego, CA, USA    (2003).  
       
\bibitem{Fer}
        F.G.S.L.\ Brand\~ao and R.O.\ Vianna,
        Phys.\ Rev.\ Lett.\ {\bf 93}, 220503 (2004).

\bibitem{Ox}
	L.M.\ Ioannou and B.C.\ Travaglione,
	Phys.\ Rev.\ A {\bf 73}, 052314 (2006).


\bibitem{Jaynes}
	R.\ Horodecki, M.\ Horodecki, and
	P.\ Horodecki, Phys.\ Rev.\ A {\bf 59}, 1799 (1999).
	
\bibitem{FM}
	F.\ Verstraete and M.M.\ Wolf, 
	Phys.\ Rev.\ Lett.\ {\bf 89}, 170401 (2002).

\bibitem{Blume}
	R.\ Blume-Kohout and P.\ Hayden,
	quant-ph/0603116.
	
\bibitem{CV1}
	L.-M.\ Duan, G.\ Giedke, J.I.\ Cirac, and P.\ Zoller, 
	Phys.\ Rev.\ Lett.\ {\bf 84}, 2722 (2000).
	
\bibitem{CV2}
	P.\ Hyllus and J.\ Eisert,
	New J.\ Phys.\ {\bf 8}, 51 (2006).
	
\bibitem{CV3}
	G.\ Giedke, M.M.\ Wolf, O.\ Krueger, 
	R.F.\ Werner, and J.I.\ Cirac, 
	Phys.\ Rev.\ Lett.\ {\bf 91}, 107901 (2003).
	
\bibitem{CV4}
	E.\ Shchukin and W.\ Vogel, 
	Phys.\ Rev.\ Lett.\ {\bf 95}, 230502 (2005).
	
\bibitem{Multi1}  
        W.\ D\"ur and J.I.\ Cirac,
        Phys.\ Rev.\ A {\bf 61}, 042314 (2000).

\bibitem{Multi2}
	 W.\ D\"ur, J.I.\ Cirac, and R.\ Tarrach,
	Phys.\ Rev.\ Lett.\ {\bf 83}, 3562 (1999).

\bibitem{Multi3}
	J.\ Eisert and D.\ Gross, 
	 {\it Multiparticle entanglement}, in
	 {\it Lectures on quantum information}, 
	 D.\ Bruss and G.\ Leuchs Eds.\ (VCH, 
	 Weinheim, 2006);
	quant-ph/0505149.
		
\bibitem{AcinThreeQubits}
        A.\ Acin, D.\ Bruss, M.\ Lewenstein, and A.\ Sanpera,
        Phys.\ Rev.\ Lett.\ {\bf 87}, 040401 (2001).
     
	
%\bibitem{Vogel} 
%	E.\ Shchukin and W.\ Vogel, 
%	Phys.\ Rev.\ Lett.\ {\bf 95}, 230502 (2005).


\bibitem{Mintert}	
	F.\ Mintert and  A.\ Buchleitner,
	quant-ph/0605250.


\bibitem{Aolita}
	L.\ Aolita and F.\ Mintert,
 	Phys.\ Rev.\ Lett.\ \textbf{97}, 50501 (2006).
	
\bibitem{Horodecki}
 	R.\ Augusiak, P.\ Horodecki, and M.\ Demianowicz,
	quant-ph/0604109.

\bibitem{Carteret}
	H.A. Carteret, 	quant-ph/0309212.
	
\bibitem{Walborn} 
  	S.P.\ Walborn, P.H.\ Souto Ribeiro, 
  	L.\ Davidovich, F.\ Mintert, and A.\ Buchleitner, 	
	Nature \textbf{440}, 1022 (2006).
	
\bibitem{Neg}
    K.\ Zyczkowski, P.\ Horodecki, A.\ Sanpera, and M.\ Lewenstein,
    Phys.\ Rev.\ A {\bf 58}, 883 (1998).
    
\bibitem{Compare}    
    J.\ Eisert and M.B.\ Plenio, 
    J.\ Mod.\ Opt.\ {\bf 46}, 145 (1999).

\bibitem{Monotone}    
    J.\ Eisert (PhD thesis, Potsdam, February 2001); see also
   quant-ph/0610253.

\bibitem{VidalWerner}    
    G.\ Vidal and R.F.\ Werner, 
    Phys.\ Rev.\ A {\bf 65}, 032314 (2002).    

\bibitem{Plenio}
	M.B.\ Plenio, 
	Phys.\ Rev.\ Lett.\ {\bf 95}, 090503 (2005).

\bibitem{Cost}
	K.\ Audenaert, M.B.\ Plenio, and J.\ Eisert,
	 Phys.\ Rev.\ Lett.\ {\bf 90}, 027901 (2003).

\bibitem{Convex}
	S.\ Boyd and L.\ Vandenberghe, {\it Convex optimization}
	(Cambridge University Press, Cambridge, 2004).


\bibitem{Koen}
	 K.M.R.\ Audenaert and S.L.\ Braunstein,
	 Comm.\ Math.\ Phys.\ {\bf 246},  443 (2004).

\bibitem{Reinhard}
	The conjugate function has -- independently
	of Ref.\ \cite{Koen} -- also been discussed in the
	context of the additivity of the entanglement
	of formation by R.F.\ Werner \cite{Priv}. 
		
\bibitem{Priv}
	R.F.\ Werner, private communication (2003).

\bibitem{Lasserre}
	J.B.\ Lasserre, SIAM J.\ Optimization
	{\bf 11}, 796 (2001).
	
\bibitem{Breuer}
	H.P.\ Breuer, 
	J.\ Phys.\ A {\bf  39}, 11847 (2006).
	
\bibitem{Chen}
	K.\ Chen, S.\ Albeverio, and S.-M.\ Fei, 
	Phys.\ Rev.\ Lett.\ 
	{\bf 95}, 040504 (2005).

 \bibitem{Vollbrecht}
 	K.G.H.\ Vollbrecht and R.F.\ Werner,
	Phys.\ Rev.\ A {\bf 64}, 062307 (2001). 

\bibitem{Durkin}    
    G.A.\ Durkin, C.\ Simon, J.\ Eisert, and D.\ Bouwmeester,
    Phys.\ Rev.\ A {\bf 70}, 062305 (2004).
        
\bibitem{OldSymm}
	 J.\ Eisert, T.\ Felbinger, P.\ Papadopoulos, 
	 M.B.\ Plenio, and M.\ Wilkens,
	 Phys.\ Rev.\ Lett.\ {\bf 84}, 1611 (2000).
	       
\bibitem{Caves}        
       K.K.\  Manne and C.M.\ Caves, quant-ph/0506151.
        
\bibitem{BreuerSymmetric}
	H.-P.\ Breuer, quant-ph/0506224.

\bibitem{Wootters}
  	W.K.\ Wootters,
  	Phys.\ Rev.\ Lett. \textbf{80}, 2245 (1998).
  	
	
\bibitem{VerstraetePhD} 
  F.\ Verstraete (PhD thesis, 
  Katholieke Universiteit Leuven, October 2002).

\bibitem{Alg}
	For example, the function {\tt NMinimize} in {\it Mathematica} 
	would be suitable.
	
  \bibitem{Geometric}
	T.-C.\ Wei and P.M.\ Goldbart, 
	Phys.\ Rev.\ A {\bf 68}, 042307 (2003). 		

\bibitem{Global}
        D.A.\ Meyer and N.R.\  Wallach,
        J.\ Math.\ Phys.\
        {\bf 43}, 4273 (2002).

\bibitem{PlenioMeasures}
	M.B.\ Plenio and S.\ Virmani,
	Quant.\ Inf.\ Comp.\ {\bf 7}, 1 (2007).

\bibitem{VidalTarrach}
  G.\ Vidal and R.\ Tarrach,
  Phys.\ Rev.\ A \textbf{59}, 141 (1999).

\bibitem{Steiner}
  M.\ Steiner,
  Phys.\ Rev.\ A \textbf{67}, 054305 (2003).


\bibitem{Brandao2}
  F.G.S.L.\ Brand\~ao,
  quant-ph/0510078.  

\bibitem{Brandao1}
  F.G.S.L.\ Brand\~ao,
  Phys.\ Rev.\ A \textbf{72}, 022310 (2005).
  
\bibitem{Cavalcanti}
  D.\ Cavalcanti and M.O.\ Terra Cunha, 
  Appl.\ Phys.\ Lett.\ {\bf 89}, 084102 (2006).
  
\bibitem{Karnas}
  S.\ Karnas and M.\ Lewenstein, 
  J.\ Phys.\ A \textbf{34}, 6919 (2001).  
  
  \bibitem{Rains}
  E.M.\ Rains, 
  IEEE Transactions on Information Theory \textbf{47}(7), 2921
  (2001). 

\bibitem{Zeil}
        D.\ Bouwmeester, J.-W.\ Pan, M.\ Daniell,
        H.\ Weinfurter, and A.\ Zeilinger,
        Phys.\ Rev.\ Lett.\ {\bf 82}, 1345 (1999).

\bibitem{Blatt}
        C.F.\ Roos,
        M.\ Riebe,
        H.\ H{\"a}ffner, W.\ H{\"a}nsel,
        J.\ Benhelm,
        G.P.T.\ Lancaster,
        C.\ Becher,
        F.\ Schmidt-Kaler, and
        R.\ Blatt, Science {\bf 304}, 1479 (2004).
 
\bibitem{ZeilingerCluster}
         P.\ Walther, K.J.\ Resch, T.\ Rudolph, E.\ Schenck, H.\ Weinfurter,
         V.\ Vedral, M.\ Aspelmeyer, and A.\ Zeilinger,
        Nature {\bf 434}, 169 (2005).

\bibitem{GS}
	D.\ Schlingemann and R.F.\ Werner,
	Phys.\ Rev.\ A {\bf 65}, 012308 (2002).
	
\bibitem{GS2}
	R.\ Raussendorf, D.E.\ Browne, and H.J.\ Briegel,
	Phys.\ Rev.\ A {\bf 68}, 022312 (2003).

\bibitem{GS3}
        M.\ Hein, J.\ Eisert, and H.J.\ Briegel,
        Phys.\ Rev.\ A {\bf 69}, 062311 (2004).

\bibitem{GS4}
	 M.\ van den Nest, J.\ Dehaene, and B.\ De Moor,
	Phys.\ Rev.\ A {\bf 70}, 034302 (2004).
	
\bibitem{GS5}		
	G. Toth and O.\ G{\"u}hne,
	Phys.\ Rev.\ A {\bf 72}, 022340 (2005).
		
\bibitem{BriegelPersistentEntanglement}
        H.J.\ Briegel and R.\ Raussendorf,
        Phys.\ Rev.\ Lett.\ {\bf 86}, 910 (2001).
	
\bibitem{Martin}
	 K.M.R.\ Audenaert and M.B.\ Plenio,
	 New J.\ Phys.\ {\bf 8}, 266 (2006).
	
\end{thebibliography}
\end{document}